\documentclass[11pt,a4paper]{article}
\usepackage[margin=2.5cm]{geometry}
\usepackage{amsmath,amssymb,amsthm}
\usepackage{booktabs,tabularx,array}
\usepackage[authoryear,round]{natbib}

\AtBeginDocument{%
  \setlength\abovedisplayskip{6pt plus 2pt minus 2pt}%
  \setlength\belowdisplayskip{6pt plus 2pt minus 2pt}}
\usepackage[colorlinks=true,linkcolor=blue,citecolor=blue,urlcolor=blue]{hyperref}
\usepackage{microtype}

\newtheorem{proposition}{Proposition}
\newtheorem{principle}{Principle}
\theoremstyle{remark}

\newcommand{\VTT}{\mathrm{VTT}}
\newcommand{\WTP}{\mathrm{WTP}}
\newcommand{\inc}{\mathrm{inc}}
\newcommand{\TT}{\mathit{tt}}
\newcommand{\TC}{\mathit{tc}}
\newcommand{\CH}{\mathit{ch}}

\title{Willingness-to-pay space and preference space:\\ a unified treatment of equivalence, heterogeneity and interpretation}
\author{Stephane Hess\thanks{Choice Modelling Centre \& Institute for Transport Studies, University of Leeds, s.hess@leeds.ac.uk}}
\date{\number\day\space\ifcase\month\or January\or February\or March\or April\or May\or June\or July\or August\or September\or October\or November\or December\fi\space\number\year}

\begin{document}
\maketitle

\begin{abstract}
\noindent A growing number of choice modelling studies estimate their models ``in willingness-to-pay (WTP) space'', yet often treat this as a different model that fits better (or worse) than ``preference space'', and misread what its parameters represent. This note brings together and makes precise insights that have mostly been stated only in scattered places, organising them around a single condition for two specifications to be the same model. WTP space is a reparameterisation, and with non-random coefficients the two spaces give identical likelihoods, WTP estimates and standard errors. With deterministic heterogeneity in cost sensitivity, or non-linearity, WTP must still be computed as a ratio of partial derivatives, although a simple normalisation recovers a directly estimated WTP at a reference point. With random heterogeneity, latent class models are equivalent in the two spaces, while other mixtures are equivalent only when the implied distributions match, as with full-covariance lognormals, and independence is not a neutral assumption.
\end{abstract}

\section{Introduction}

Since \citet{TrainWeeks2005}, a growing number of studies specify discrete choice models ``in WTP space'', where the marginal utilities of non-cost attributes are written as the product of a cost coefficient and a monetary valuation, rather than ``in preference space'', where each attribute has its own marginal utility \citep[see also][]{SonnierEtAl2007,ScarpaThieneTrain2008}. The approach is useful, especially with random coefficients, because it lets the analyst place distributional assumptions directly on WTP and avoid the problems that arise from ratios of random coefficients \citep{DalyHessTrain2012}.

Its use is, however, surrounded by persistent confusion about what WTP space is, how it relates to preference space and what its parameters represent, even though the key conceptual points have been made before. \citet{TrainWeeks2005} and \citet{ScarpaThieneTrain2008} state that the two approaches are formally equivalent, that they differ in practice only because convenient distributions mean different things in the two spaces, and that independence restrictions are asymmetric between them. For models with non-random coefficients, \citet{DalyHessDeJong2012} and \citet{DalyHessOrtuzar2023} show that WTP computed from preference-space estimates, and its Delta method standard error, are the maximum likelihood estimate and its asymptotic standard error, and that the two spaces give the same results in practice. \citet{HessRose2012} conclude that a better fit in WTP space ``is simply a reflection that the resulting distributional assumptions better match the data'', and \citet{HessTrain2017} show that WTP space with full covariance allows for all sources of correlation.

These points have nevertheless not been absorbed. Applied papers continue to present the two ``spaces'' as rival models and to credit differences in fit to the space itself \citep[e.g.][]{BalcombeEtAl2009,RoseMasiero2010,OwusuCoffieEtAl2016,Chen2026}. Models are described as having the cost coefficient ``normalised to one'', and the cost coefficient in WTP space is routinely labelled ``the scale parameter'', which has led to claims that WTP space cannot accommodate differences in price sensitivity across groups \citep{Helveston2023}. From a more technical point of view, the loss of global concavity of the likelihood is rarely discussed, beyond the greater sensitivity to starting values noted by \citet{Helveston2023}, and independence between coefficients is treated as a neutral default. Part of the problem is that, beyond the case without random heterogeneity, the earlier insights are mostly stated verbally and in passing, in papers whose main focus lies elsewhere.

This note therefore aims to clarify and unify. We organise the discussion around a single condition for two specifications to be the same model (Principle~\ref{pr:equiv}), restate the established points precisely, including the equivalence of the two spaces with non-random coefficients (Proposition~\ref{prop:fixed}), and add points that we have not seen stated, on deterministic heterogeneity (Section~\ref{sec:det}) and on a reference-point normalisation for non-linear cost (Section~\ref{sec:nonlin}). For random heterogeneity, we set out when discrete and continuous mixtures are equivalent in the two spaces, including exact equivalence for latent class models and for continuous mixtures with unrestricted correlation of the log-coefficients, and the correlations that independence implies in each space (Section~\ref{sec:random}).

\section{WTP space as a reparameterisation}\label{sec:reparam}

Throughout, we use travel time ($\TT$) and travel cost ($\TC$) as the running example, with the value of travel time ($\VTT$) as the WTP measure and, where needed, a second non-cost attribute, the number of interchanges ($\CH$). The transport context is used solely for illustration. 

Consider the preference-space utility of alternative $j$ for person $n$:
\begin{equation}\label{eq:pref}
V_{jn} = \delta_j + \beta_{\TC} \TC_{jn} + \beta_{\TT} \TT_{jn} + \beta_{\CH} \CH_{jn},
\end{equation}
where the WTP for a non-cost attribute $x_k$ (here $\TT$ or $\CH$) is given by the marginal rate of substitution (MRS) between the attribute and cost, i.e.\ the ratio of partial derivatives,
\begin{equation}\label{eq:mrs}
\WTP_{k,n} = \frac{\partial V_{jn}/\partial x_{k,jn}}{\partial V_{jn}/\partial \TC_{jn}},
\end{equation}
so that $\VTT = \beta_{\TT}/\beta_{\TC}$ and $\WTP_{\CH} = \beta_{\CH}/\beta_{\TC}$. With this definition, the WTP is positive for an attribute that, like cost, reduces utility, and is then the WTP for a reduction in that attribute. For an attribute that increases utility, the ratio is negative, and the WTP for an increase is minus the ratio.

A helpful way to see what a reparameterisation involves is to think about units. If time entered the utility function in hours rather than minutes, and cost in pence rather than pounds, we could write
$V_{jn} = \delta_j + \beta_{\TC}' (100\,\TC_{jn}) + \beta_{\TT}' (\TT_{jn}/60) + \beta_{\CH} \CH_{jn}$,
with the attributes $\TT_{jn}$ and $\TC_{jn}$ themselves still measured in minutes and pounds.
The model is unchanged and its likelihood is identical, with $\beta_{\TT}' = 60\beta_{\TT}$ and $\beta_{\TC}' = \beta_{\TC}/100$. Nobody would thus call this a different model. 

WTP space applies the same logic but changes the \emph{numeraire} of utility rather than the units of an attribute. Writing $\beta_{\TT} = \beta_{\TC}\beta_{v,\TT}$ and $\beta_{\CH} = \beta_{\TC}\beta_{v,\CH}$ gives
\begin{align}
V_{jn} &= \delta_j + \beta_{\TC}\,\TC_{jn} + \beta_{\TC}\beta_{v,\TT}\,\TT_{jn} + \beta_{\TC}\beta_{v,\CH}\,\CH_{jn}\\
&= \delta_j + \beta_{\TC}\left(\TC_{jn} + \beta_{v,\TT}\, \TT_{jn} + \beta_{v,\CH}\, \CH_{jn}\right). \label{eq:wtp}
\end{align}
The ratio of partial derivatives gives \mbox{$\VTT = (\partial V_{jn}/\partial \TT_{jn})/(\partial V_{jn}/\partial \TC_{jn}) = \beta_{\TC}\beta_{v,\TT}/\beta_{\TC} = \beta_{v,\TT}$}, confirming that $\beta_{v,\TT}$ is a directly estimated VTT, and the same applies to $\beta_{v,\CH}$ as the WTP for $\CH$. In what follows, we refer to the expression in brackets in \eqref{eq:wtp}, which is utility expressed in money, as \emph{monetised utility}. Multiplying it by the cost coefficient $\beta_{\TC}$, i.e.\ minus the marginal utility of income, converts it back into utility.

Both the change of units and the move to WTP space are instances of a general condition, which we use throughout.

\begin{principle}[Equivalence]\label{pr:equiv}
A preference-space and a WTP-space specification with the same specification of the error term are the same model if and only if, up to the normalisations needed for identification, they imply for every person the same alternative specific constants and the same partial derivatives of utility with respect to every attribute, measured as in the data. With random heterogeneity, the condition is that the joint distribution of these constants and partial derivatives across people is the same.
\end{principle}

For the change of units, $\partial V_{jn}/\partial\TC_{jn} = 100\,\beta_{\TC}' = \beta_{\TC}$ and $\partial V_{jn}/\partial\TT_{jn} = \beta_{\TT}'/60 = \beta_{\TT}$, so the two specifications are the same model. For \eqref{eq:pref} and \eqref{eq:wtp}, the partial derivative with respect to cost is $\beta_{\TC}$ in both, while the partial derivative with respect to time is $\beta_{\TT}$ in \eqref{eq:pref} and $\beta_{\TC}\beta_{v,\TT}$ in \eqref{eq:wtp}, and similarly for $\CH$. These are equal for $\beta_{v,\TT} = \beta_{\TT}/\beta_{\TC}$ and $\beta_{v,\CH} = \beta_{\CH}/\beta_{\TC}$, which is well defined whenever $\beta_{\TC}\neq0$. Equations~\eqref{eq:pref} and \eqref{eq:wtp} are thus two parameterisations of the same utility function, and \citet[Section 5]{DalyHessOrtuzar2023} illustrate this empirically.

As WTP measures are ratios of partial derivatives \eqref{eq:mrs}, two equivalent specifications also imply the same WTP measures. The converse does not hold, as two specifications can imply the same ratios of partial derivatives, and thus the same WTP measures, but different partial derivatives, in which case they are different models (Section~\ref{sec:det}). Table~\ref{tab:confusions} summarises common statements about WTP space that do not hold.

\begin{table}[t]
\centering
\caption{Common statements about WTP space and what actually holds.}\label{tab:confusions}
\small
\begin{tabularx}{\textwidth}{@{}>{\raggedright\arraybackslash}p{0.27\textwidth}>{\raggedright\arraybackslash}X@{}}
\toprule
\emph{Common statement} & \emph{What actually holds}\\
\midrule
\textbf{``The cost coefficient is constrained to one''} & $\beta_{\TC}$ is estimated. The ``1'' in front of $\TC_{jn}$ in \eqref{eq:wtp} is the reparameterisation itself, not a restriction.\\
\addlinespace[6pt]
\textbf{``The cost coefficient is a scale parameter''} & $\beta_{\TC}$ is the marginal utility of cost, i.e.\ minus the marginal utility of income, as in preference space, although heterogeneity placed on it acts like scale heterogeneity.\\
\addlinespace[6pt]
\textbf{``All non-cost terms must be scaled by $\boldsymbol{\beta_{\TC}}$''} & Which terms are expressed in money, e.g.\ whether $\delta_j$ is included in monetised utility, is the analyst's choice, though it has implications with random heterogeneity (Section~\ref{sec:random}).\\
\addlinespace[6pt]
\textbf{``Non-cost terms need a negative sign''} & The sign only determines whether the valuations are the WTP for a reduction or an increase in the attribute. Nothing else changes.\\
\addlinespace[6pt]
\textbf{``Cost must be the numeraire''} & Any attribute can serve as the numeraire (e.g.\ ``time space''), using the denominator of the marginal rate of substitution of interest.\\
\bottomrule
\end{tabularx}
\end{table}

Two of these confusions have identifiable origins. In a logit model, utility is identified only up to scale, and the usual normalisation fixes the scale of the error term and estimates all coefficients. An equivalent alternative is to fix the coefficient of one attribute and estimate the scale instead. When this attribute is cost, with its coefficient fixed to $-1$, the estimated scale takes the role of $-\beta_{\TC}$ and the remaining coefficients become valuations, so that the model is exactly the WTP-space model \citep[][eq.~10]{FiebigEtAl2010,HessTrain2017}. The language of a price coefficient ``normalised to one'' is, however, carried over to the usual implementation of WTP space, where the error scale is fixed and the cost coefficient is estimated\footnote{This ambiguity does not arise in software where the analyst writes the utility function directly, such as Apollo \citep{HessPalma2019}.}, not least when WTP-space models are fitted with generalised multinomial logit (G-MNL) software such as \texttt{gmnl} \citep{GuHoleKnox2013}, and WTP space is then described as restricting the price coefficient to one \citep[e.g.][]{OwusuCoffieEtAl2016,Chen2026}. The ``scale parameter'' label \citep{TrainWeeks2005,Helveston2023} arises because $\beta_{\TC}$ multiplies all monetised terms, so that any heterogeneity placed on it shifts all monetised marginal utilities in proportion, as scale heterogeneity does (Sections~\ref{sec:det} and \ref{sec:corr}). The label is misleading for two reasons. First, when the constants are not included in monetised utility, as with $\delta_j$ in \eqref{eq:wtp}, $\beta_{\TC}$ does not multiply all of utility and is thus by definition not a scale parameter. Second, heterogeneity in $\beta_{\TC}$ need not be scale heterogeneity \citep{HessRose2012,HessTrain2017}, as sensitivity to cost alone can vary in WTP space through the cost attribute in monetised utility (Section~\ref{sec:det}).

The reparameterisation does have one practical cost. Equation~\eqref{eq:pref} is linear in parameters, so the multinomial logit log-likelihood is globally concave \citep{McFadden1974}. Equation~\eqref{eq:wtp} is non-linear in parameters, so global concavity is lost. With non-random coefficients the maximum is still unique, provided the preference-space maximum likelihood estimate exists and has $\hat\beta_{\TC}\neq0$. However, the likelihood has saddle points where $\beta_{\TC}=0$, and an optimiser started with the wrong sign for $\beta_{\TC}$ can drift towards zero while the valuations grow without bound, so estimation is more sensitive to starting values \citep{Helveston2023}. Preference space thus has the advantage of avoiding a utility specification that is non-linear in parameters. With mixtures, local optima are a concern in both spaces, but the non-linearity of the WTP-space specification further amplifies the problem (Section~\ref{sec:random}).

\section{Models without random heterogeneity}\label{sec:norandom}

We first consider models without random heterogeneity. Alternative specific constants are omitted from the specifications below for brevity, as they enter outside monetised utility, as in \eqref{eq:pref} and \eqref{eq:wtp}, and do not affect the arguments.

\subsection{Equivalence of estimates and standard errors}\label{sec:fixed}

In preference space, WTP is computed after estimation as a function $g(\hat\theta)$ of the estimates, e.g.\ $g(\hat\theta) = \hat\beta_{\TT}/\hat\beta_{\TC}$, with its standard error usually obtained with the Delta method, which gives the covariance matrix of $g(\hat\theta)$ as $\nabla g\,\widehat{\mathrm{Var}}(\hat\theta)\,\nabla g^\top$, where $\nabla g=\partial g/\partial\theta^\top$ is evaluated at $\hat\theta$. In WTP space, the valuations are instead parameters of the model, and their standard errors come directly from the estimated covariance matrix of the parameters.

\begin{proposition}[\citealp{DalyHessDeJong2012,DalyHessOrtuzar2023}]\label{prop:fixed}
Let $\theta$ denote the parameters of the preference-space model and $\phi$ those of the WTP-space model. The two are related by $\phi = h(\theta)$, given in Section~\ref{sec:reparam}, with $h$ one-to-one and continuously differentiable, with an invertible Jacobian, for $\beta_{\TC}\neq 0$. For each set of preference-space coefficients, there is therefore exactly one set of WTP-space coefficients for which the partial derivatives of utility with respect to all attributes take the same values, so that Principle~\ref{pr:equiv} holds. Let $\hat\theta$ and $\hat\phi$ be the maximum likelihood estimates obtained by estimating each model separately. Then (i) the maximised log-likelihoods are identical and $\hat\phi = h(\hat\theta)$, i.e.\ the estimates obtained in WTP space are exactly the transformed preference-space estimates, and (ii) the asymptotic covariance matrix of $\hat\phi$ obtained from WTP-space estimation, whether based on the Hessian, the outer product of the gradients \citep{BHHH1974} or the robust sandwich, equals the Delta method covariance $J\,\widehat{\mathrm{Var}}(\hat\theta)\,J^\top$, with $\widehat{\mathrm{Var}}(\hat\theta)$ computed in the same way in preference space and $J=\partial h/\partial\theta^\top$ evaluated at $\hat\theta$.
\end{proposition}

Part (i) is the invariance of maximum likelihood estimation to reparameterisation. Part (ii) holds because, at the maximum, the Hessian, the outer product of the gradients and the individual score contributions all transform exactly with $J$, as shown by \citet{DalyHessDeJong2012} for the Hessian and by \citet[Appendix 1]{DalyHessOrtuzar2023} for the other estimators. Together, (i) and (ii) mean that $g(\hat\theta)$ is itself the maximum likelihood estimate of $g(\theta)$, and that its Delta method standard error is the asymptotic maximum likelihood standard error, not an approximation to it.

Three consequences follow. First, with non-random coefficients, ``obtaining standard errors directly'' is a convenience rather than an advantage of WTP space. Second, beyond small numerical differences arising from convergence tolerances and numerical or approximated second derivatives, any differences between the two spaces in applied work with non-random coefficients point to non-convergence or to different specifications, not to a property of the ``space''. Third, what \emph{does} differ is the shape of the likelihood away from the optimum, so that likelihood-ratio (profile) confidence intervals for WTP do not depend on the parameterisation, while simulation-based (Krinsky--Robb type) intervals do, and should not be used for standard errors of ratios \citep{DalyHessOrtuzar2023}\footnote{Similarly, what a $t$-ratio tests depends on the parameterisation. With a lognormal $-\beta_{\TC}$ (Section~\ref{sec:continuous}), the $t$-ratio of the location parameter of $\ln(-\beta_{\TC})$ tests whether the median of $-\beta_{\TC}$ is one, so it changes arbitrarily with the currency unit, without changing the model.}.

\subsection{Deterministic heterogeneity}\label{sec:det}

Whatever the specification, the WTP measure is \emph{always} the ratio of partial derivatives in \eqref{eq:mrs}. Throughout, a subscript $n$ denotes a value specific to person $n$, which in this section varies deterministically with characteristics of that person. Write a general WTP-space utility as $V_{jn} = \beta_{\TC,n}\left[q_{\TC,n}(\TC_{jn}) + \sum_k \beta_{v,k,n}\,q_{k,n}(x_{k,jn})\right]$, where the functions $q_{\TC,n}$ and $q_{k,n}$ allow for non-linearity and for interactions with characteristics of person $n$, and $q_{\TC,n}'$ and $q_{k,n}'$ denote their derivatives with respect to cost and attribute $k$, respectively. Then
\begin{equation}\label{eq:wtp_general}
\WTP_{k,n} = \frac{\partial V_{jn}/\partial x_{k,jn}}{\partial V_{jn}/\partial \TC_{jn}} = \frac{\beta_{\TC,n}\,\beta_{v,k,n}\,q_{k,n}'(x_{k,jn})}{\beta_{\TC,n}\,q_{\TC,n}'(\TC_{jn})} = \frac{\beta_{v,k,n}\,q_{k,n}'(x_{k,jn})}{q_{\TC,n}'(\TC_{jn})},
\end{equation}
so the estimated valuation parameter $\beta_{v,k,n}$ is the WTP only when $q_{k,n}'(x_{k,jn})/q_{\TC,n}'(\TC_{jn})=1$, which in practice means that both cost and the attribute enter monetised utility linearly and without interactions. A constant ratio other than one only changes the units in which the WTP is measured. 

In this formulation, heterogeneity can enter in two places, with different consequences. Heterogeneity in monetised utility, through $q_{\TC,n}$, $q_{k,n}$ or $\beta_{v,k,n}$, affects WTP. Heterogeneity in the cost coefficient $\beta_{\TC,n}$, which multiplies all of these terms, cancels from every WTP measure and affects only the scale of monetised utility.

\subsubsection{Heterogeneity in sensitivity to a non-cost attribute}

Let $z_n$ be a dummy variable equal to one if person $n$ is male and zero otherwise, and suppose that gender shifts the sensitivity to travel time. In preference space, this gives
\begin{gather}
V_{jn} = \beta_{\TC}\,\TC_{jn} + \left(\beta_{\TT} + \beta_{\TT,z}z_n\right)\TT_{jn} + \beta_{\CH} \CH_{jn}, \label{eq:gender_pref}\\
\VTT_n = \frac{\partial V_{jn}/\partial \TT_{jn}}{\partial V_{jn}/\partial \TC_{jn}} = \frac{\beta_{\TT} + \beta_{\TT,z}z_n}{\beta_{\TC}},\qquad \WTP_{\CH,n} = \frac{\beta_{\CH}}{\beta_{\TC}},
\end{gather}
and in WTP space
\begin{gather}
V_{jn} = \beta_{\TC}\left[\TC_{jn} + \left(\beta_{v,\TT}+\beta_{v,\TT,z}z_n\right)\TT_{jn} + \beta_{v,\CH}\CH_{jn}\right], \label{eq:gender_wtp}\\
\VTT_n = \frac{\partial V_{jn}/\partial \TT_{jn}}{\partial V_{jn}/\partial \TC_{jn}} = \frac{\beta_{\TC}\left(\beta_{v,\TT}+\beta_{v,\TT,z}z_n\right)}{\beta_{\TC}} = \beta_{v,\TT}+\beta_{v,\TT,z}z_n,\qquad \WTP_{\CH,n} = \beta_{v,\CH}.
\end{gather}
The two are reparameterisations of each other, in line with Principle~\ref{pr:equiv}, with $\beta_{v,\TT} = \beta_{\TT}/\beta_{\TC}$ and $\beta_{v,\TT,z} = \beta_{\TT,z}/\beta_{\TC}$. In terms of \eqref{eq:wtp_general}, the heterogeneity enters through the valuation parameter, while cost and time still enter monetised utility linearly, so the VTT is read directly, with $\beta_{v,\TT}$ the VTT for women, $\beta_{v,\TT}+\beta_{v,\TT,z}$ the VTT for men, and $\beta_{v,\TT,z}$ the difference between them, while the WTP for $\CH$ is unaffected.

\subsubsection{Heterogeneity in cost sensitivity}

Now let income affect cost sensitivity, with $\rho_n = (\inc_n/\overline{\inc})^{\lambda}$, where $\inc_n$ is the income for person $n$, $\overline{\inc}$ is the mean income in the sample, and $\lambda$ is an estimated income elasticity:
\begin{gather}
V_{jn} = \beta_{\TC}\,\rho_n\,\TC_{jn} + \beta_{\TT} \TT_{jn} + \beta_{\CH} \CH_{jn}, \label{eq:inc_pref}\\
\VTT_n = \frac{\partial V_{jn}/\partial \TT_{jn}}{\partial V_{jn}/\partial \TC_{jn}} = \frac{\beta_{\TT}}{\beta_{\TC}\rho_n},\qquad \WTP_{\CH,n} = \frac{\beta_{\CH}}{\beta_{\TC}\rho_n}.
\end{gather}
Income changes \emph{every} WTP measure proportionally, with a common elasticity $-\lambda$. The correct WTP-space counterpart puts the interaction on cost within monetised utility:
\begin{gather}
\begin{aligned}
V_{jn} &= \beta_{\TC}\left[\rho_n \TC_{jn} + \beta_{v,\TT}\TT_{jn} + \beta_{v,\CH}\CH_{jn}\right]\\
&= (\beta_{\TC}\rho_n)\left[\TC_{jn} + \frac{\beta_{v,\TT}}{\rho_n}\TT_{jn} + \frac{\beta_{v,\CH}}{\rho_n}\CH_{jn}\right],
\end{aligned} \label{eq:inc_correct}\\
\VTT_n = \frac{\partial V_{jn}/\partial \TT_{jn}}{\partial V_{jn}/\partial \TC_{jn}} = \frac{\beta_{\TC}\beta_{v,\TT}}{\beta_{\TC}\rho_n} = \frac{\beta_{v,\TT}}{\rho_n},\qquad \WTP_{\CH,n} = \frac{\beta_{v,\CH}}{\rho_n}.
\end{gather}
The second line shows that income operates only through the marginal utility of income, $\beta_{\TC}\rho_n$, with the bracket again in money units, while the marginal utilities of time and interchanges, $\beta_{\TC}\beta_{v,\TT}$ and $\beta_{\TC}\beta_{v,\CH}$, do not vary with income. Equation~\eqref{eq:inc_correct} is a reparameterisation of \eqref{eq:inc_pref}, so the two are identical by Principle~\ref{pr:equiv}, and Proposition~\ref{prop:fixed} applies. However, $\beta_{v,\TT}$ and $\beta_{v,\CH}$ themselves are now WTP measures only for a person with mean income, and $\VTT_n$ thus has to be computed explicitly using \eqref{eq:mrs}.

This example also shows that the claim that WTP space cannot accommodate differences in price sensitivity across groups, because ``only one scale parameter is estimated'' \citep{Helveston2023}, is mistaken: cost sensitivity can be made to vary with covariates exactly as in preference space, by interacting the cost attribute in monetised utility, as in \eqref{eq:inc_correct}. By contrast, interacting the coefficient $\beta_{\TC}$ that multiplies monetised utility, as the ``scale parameter'' view would suggest, gives a different specification:
\begin{equation}\label{eq:incB}
V_{jn} = \beta_{\TC}\rho_n\left[\TC_{jn} + \beta_{v,\TT}\TT_{jn} + \beta_{v,\CH}\CH_{jn}\right],\qquad \VTT_n = \beta_{v,\TT},\qquad \WTP_{\CH,n} = \beta_{v,\CH}.
\end{equation}
Here income is a pure scale effect. Unlike in \eqref{eq:inc_correct}, it changes the marginal utilities of time and interchanges in the same proportion as the marginal utility of income, and thus cancels from every WTP measure, so it cannot represent an effect of income on WTP.

It may also be tempting to place the income effect on the valuations instead of on cost:
\begin{equation}\label{eq:incD}
V_{jn} = \beta_{\TC}\left[\TC_{jn} + \frac{\beta_{v,\TT}\TT_{jn} + \beta_{v,\CH}\CH_{jn}}{\rho_n}\right],\qquad \VTT_n = \frac{\beta_{v,\TT}}{\rho_n},\qquad \WTP_{\CH,n} = \frac{\beta_{v,\CH}}{\rho_n}.
\end{equation}
Income then affects all WTP measures in the same proportion, and since the utility in \eqref{eq:inc_correct} is that in \eqref{eq:incD} multiplied by $\rho_n$, the two give the same WTP measures at the same parameter values. They are nevertheless different models (Principle~\ref{pr:equiv}), as $\rho_n$ changes the partial derivatives of utility, and hence the choice probabilities, differently for different people, so that on the same data they will in general fit differently and give different WTP estimates\footnote{This assumes the same error variance for all individuals, as otherwise, with the scale of utility relative to the error term varying in proportion to $\rho_n$, \eqref{eq:inc_correct} and \eqref{eq:incD} could not be distinguished.}. Specification \eqref{eq:incD} has a preference-space counterpart, in which income scales all non-cost coefficients, and in either space only \eqref{eq:inc_correct}, like \eqref{eq:inc_pref}, represents an income effect operating through the marginal utility of income, as is also the case when covariates are interacted with each valuation parameter separately \citep[e.g.][]{LadenburgSkotte2022}.

\subsection{Non-linearity}\label{sec:nonlin}

\citet{HessTrain2017} note that WTP space ``is not useful when the price does not enter utility linearly, since, with nonlinear price effects, WTP is not simply the ratio of the attribute coefficient to the coefficient of the price variable''. The same issue as in Section~\ref{sec:det} arises, but it can be resolved. Take the common log-cost specification,
\begin{gather}
V_{jn} = \beta_{\TC} \ln \TC_{jn} + \beta_{\TT} \TT_{jn} + \beta_{\CH} \CH_{jn}, \label{eq:log_pref}\\
\VTT_{jn} = \frac{\partial V_{jn}/\partial \TT_{jn}}{\partial V_{jn}/\partial \TC_{jn}} = \frac{\beta_{\TT}}{\beta_{\TC}/\TC_{jn}} = \frac{\beta_{\TT}}{\beta_{\TC}}\,\TC_{jn},\qquad \WTP_{\CH,jn} = \frac{\beta_{\CH}}{\beta_{\TC}}\,\TC_{jn}.
\end{gather}
Unlike in Section~\ref{sec:det}, the WTP measures now vary not only across people but also across alternatives, as they depend on the cost of the alternative, hence the subscript $jn$. The ``natural'' WTP-space version is
\begin{equation}\label{eq:log_wtp}
V_{jn} = \beta_{\TC}\left[\ln \TC_{jn} + \beta_{v,\TT}\TT_{jn} + \beta_{v,\CH}\CH_{jn}\right].
\end{equation}
This is again equivalent, but the bracket is now in units of $\ln\TC$ rather than money, so that $\beta_{\TC}$ is no longer minus the marginal utility of income, which now varies with cost, and $\beta_{v,\TT}$ and $\beta_{v,\CH}$ are WTP measures only at $\TC_{jn}=1$, a point that depends on the currency unit and may well lie outside the range of the data. Both WTP measures vary with $\TC_{jn}$ and must again be obtained from \eqref{eq:mrs}, giving
\begin{equation}
\VTT_{jn} = \frac{\partial V_{jn}/\partial \TT_{jn}}{\partial V_{jn}/\partial \TC_{jn}} = \frac{\beta_{\TC}\beta_{v,\TT}}{\beta_{\TC}/\TC_{jn}} = \beta_{v,\TT}\,\TC_{jn},\qquad \WTP_{\CH,jn} = \beta_{v,\CH}\,\TC_{jn}.
\end{equation}

If the analyst wants a parameter that is the WTP at a meaningful point, the non-linear transformation can be normalised so that its derivative equals one at a chosen reference value $\TC^\ast$ (e.g.\ the sample mean cost), rather than at $\TC_{jn}=1$:
\begin{gather}
V_{jn} = \beta_{\TC}^\ast\left[\TC^\ast\ln \TC_{jn} + \beta^\ast_{v,\TT}\TT_{jn} + \beta^\ast_{v,\CH}\CH_{jn}\right], \label{eq:log_norm}\\
\VTT_{jn} = \beta^\ast_{v,\TT}\,\frac{\TC_{jn}}{\TC^\ast},\qquad \WTP_{\CH,jn} = \beta^\ast_{v,\CH}\,\frac{\TC_{jn}}{\TC^\ast}.
\end{gather}
Now $\beta^\ast_{v,\TT}$ and $\beta^\ast_{v,\CH}$ are the $\VTT$ and $\WTP_{\CH}$ at $\TC_{jn}=\TC^\ast$, and are estimated, with their standard errors, directly. Equivalently, the bracket is in money units at $\TC^\ast$, where $\beta^\ast_{\TC}$ is minus the marginal utility of income. Equation~\eqref{eq:log_norm} is again just a reparameterisation of \eqref{eq:log_pref}, in line with Principle~\ref{pr:equiv}, with $\beta^\ast_{\TC} = \beta_{\TC}/\TC^\ast$, $\beta^\ast_{v,\TT} = \TC^\ast\beta_{\TT}/\beta_{\TC}$ and $\beta^\ast_{v,\CH} = \TC^\ast\beta_{\CH}/\beta_{\TC}$. As $\TC^\ast$ is chosen by the analyst rather than estimated, the fit, the $\VTT_{jn}$ and $\WTP_{\CH,jn}$ at every point and, by Proposition~\ref{prop:fixed}, the standard errors of any derived quantity are unchanged, and $\TC^\ast$ only determines the point to which the reported valuation parameters refer\footnote{This also holds with random coefficients (Section~\ref{sec:random}), provided the parameters of the distribution, including any bounds, are estimated, as rescaling a random coefficient keeps it in the same family (for a lognormal, only the location shifts).}. The same device works for any differentiable transformation $q_{\TC}(\TC)$, such as a Box--Cox transformation, using $q_{\TC}(\TC)/q_{\TC}'(\TC^\ast)$, and for a non-linear transformation $q_{\TT}(\TT)$ of time, using $q_{\TT}(\TT)/q_{\TT}'(\TT^\ast)$. The valuation parameters are then the MRS at the reference point, which is $(\TC^\ast,\TT^\ast)$ for the $\VTT$, while $\WTP_{\CH}$ requires only the cost normalisation. This is an application of \eqref{eq:wtp_general}: a valuation parameter is the WTP only where the derivatives of the transformations in monetised utility equal one, and the normalisation simply chooses that point. This generalises the common practice of defining valuations at reference values of covariates and trip characteristics, as in large-scale value-of-time studies \citep[e.g.][]{HessEtAl2017}, to non-linear transformations of the attributes.

Finally, the MRS, and hence any valuation parameter, is a marginal value, i.e.\ the WTP per unit for a very small change. The WTP for a finite change is instead the amount of money that leaves utility unchanged, and with non-linear cost this is not simply the MRS multiplied by the size of the change, in either space. In \eqref{eq:log_norm}, the WTP $w$ for a saving of $\Delta\TT$ minutes solves $\TC^\ast\ln(\TC_{jn}+w) - \TC^\ast\ln \TC_{jn} = \beta^\ast_{v,\TT}\Delta\TT$, i.e.\ $w = \TC_{jn}\left[\exp(\beta^\ast_{v,\TT}\Delta\TT/\TC^\ast)-1\right]$. With $\TC_{jn}=\TC^\ast=10$ and $\beta^\ast_{v,\TT}=0.2$ per minute, for example, a saving of 30 minutes gives $w\approx 8.2$, against $30\times0.2=6$ from the MRS. The same applies to welfare measures more generally. With non-linear cost, the marginal utility of income varies with the cost of the alternative, so that the usual logsum measure of consumer surplus \citep{SmallRosen1981} no longer applies, and compensating variation requires the more general approach of \citet{DagsvikKarlstrom2005}.

\subsection{Implications}\label{sec:implications}

The convenience of WTP space, namely that the estimated parameters are the WTP measures, holds only when cost and all attributes enter monetised utility linearly, with any deterministic heterogeneity entering additively through the valuation parameters. Otherwise, WTP must again be computed as a ratio of partial derivatives, with Delta method standard errors, exactly as in preference space. Since realistic specifications almost always include some deterministic heterogeneity or non-linearity, not least in cost, this convenience is limited in practice.

What can be retained is a single directly estimated value at a reference point, by centring covariates, as with mean income in \eqref{eq:inc_correct}, or by normalising non-linear transformations, as in \eqref{eq:log_norm}. This gives the WTP for a person with the reference characteristics, facing the reference attribute levels. The mean WTP in the sample is instead obtained by integrating the WTP over the distribution of the covariates and attribute levels, and in general differs from the value at the reference point unless the WTP is linear in these. Computing it, like any richer presentation of results, again requires ratios of partial derivatives, so that WTP space offers no advantage here.

\section{Models with random heterogeneity}\label{sec:random}

With random heterogeneity the two spaces are no longer automatically equivalent, and this is where the claims of ``better fit'' come from. \citet{TrainWeeks2005}, \citet{HensherGreene2011}, \citet{HoleKolstad2012} and \citet{Chen2026} find that preference space fits better (although only marginally for the best-fitting models in the latter two), while \citet{ScarpaThieneTrain2008}, \citet{BalcombeEtAl2009}, \citet{RoseMasiero2010} and \citet{OwusuCoffieEtAl2016} find that WTP space fits better (although in \citealt{RoseMasiero2010} only the WTP-space models allow for heterogeneity in cost sensitivity). \citet{SonnierEtAl2007} find mixed results in-sample, but better out-of-sample prediction in WTP space. Several of these papers present the result as evidence for one ``space'' over the other. 

The reason for the differences is not that one space is a better model, but that the analyst specifies distributions for different quantities \citep{TrainWeeks2005,HessRose2012}. What has received less attention is that whether this matters depends on the type of mixing distribution.

\subsection{General setting}\label{sec:general}

Let $\theta = (\beta_{\TC},\beta_{\TT},\beta_{\CH})$ denote the preference-space coefficients and $\phi = h(\theta) = (\beta_{\TC},\beta_{v,\TT},\beta_{v,\CH})$ their WTP-space counterparts, with $\beta_{v,\TT} = \beta_{\TT}/\beta_{\TC}$ and $\beta_{v,\CH} = \beta_{\CH}/\beta_{\TC}$. Alternative specific constants included in monetised utility are valuations like any other and are included in $\theta$ and $\phi$, while constants outside monetised utility, as $\delta_j$ in \eqref{eq:pref}, are specified in the same way in both spaces and can be left out. Heterogeneity is represented by a mixing distribution $G(\theta\mid\Omega)$, where $\Omega$ contains the parameters of the distribution, such as means and covariances, and we write $P_{Y_{nt}}(\theta)$ for the probability, conditional on $\theta$, of $Y_{nt}$, the alternative chosen by person $n$ in choice situation $t=1,\dots,T_n$. We use the standard specification for repeated choices of \citet{ReveltTrain1998}, in which the coefficients vary across people but are constant across the choices made by the same person. As before, the subscript $n$ denotes the value for person $n$, and is dropped when referring to the distribution across people. The likelihood contribution of person $n$ is
\begin{align}
\text{general:}\qquad & L_n(\Omega) = E_G\Big[\textstyle\prod_{t=1}^{T_n} P_{Y_{nt}}(\theta)\Big], \label{eq:mix_gen}\\
\text{continuous mixture:}\qquad & L_n(\Omega) = \int \textstyle\prod_{t=1}^{T_n} P_{Y_{nt}}(\theta)\, f(\theta\mid\Omega)\,d\theta, \label{eq:mix_cont}\\
\text{discrete mixture:}\qquad & L_n(\Omega) = \textstyle\sum_{c=1}^{C} \pi_{nc} \prod_{t=1}^{T_n} P_{Y_{nt}}(\theta_c), \label{eq:mix_disc}
\end{align}
where $f$ is the density of $G$ and, in the discrete case, $G$ places probability $\pi_{nc}$ on support point $\theta_c$, with the probabilities possibly depending on covariates of person $n$. Writing the same model in WTP space means specifying a mixing distribution $G^\phi(\phi\mid\Omega^\phi)$ for $\phi$, and since $P_{Y_{nt}}(\theta) = P_{Y_{nt}}(h^{-1}(\phi))$, a preference-space model with $G$ and a WTP-space model with $G^\phi$ are the \emph{same} model if and only if $G^\phi$ is the distribution of $h(\theta)$ under $G$. This is Principle~\ref{pr:equiv} applied to mixtures. The model is defined by the mixing distribution, and the ``space'' only determines the coordinates in which the analyst chooses to specify it. In WTP space, the analyst also decides which terms are multiplied by $\beta_{\TC,n}$. A constant $\kappa_j$ inside monetised utility, which enters utility as $\beta_{\TC,n}\kappa_j$, and a constant $\delta_j$ outside it are reparameterisations of each other ($\delta_j = \beta_{\TC}\kappa_j$) with a non-random cost coefficient, but different models with a random one, as the effect of $\kappa_j$ then varies across people with $\beta_{\TC,n}$.

\subsection{Discrete mixtures}\label{sec:discrete}

Two different specifications are used in practice, and they behave differently. In a latent class model, each class is a full model with its own non-random coefficients, while in discrete mixtures with separate support for each coefficient, as in continuous mixtures (Section~\ref{sec:continuous}), it is the coefficients themselves that are given a distribution, and equivalence then depends on whether the distributions specified in the two spaces match.

\subsubsection{Latent class models}\label{sec:lc}

In a latent class model \citep[e.g.][]{GreeneHensher2003}, each support point is a class with its own full coefficient vector $\theta_c = (\beta_{\TC,c},\beta_{\TT,c},\beta_{\CH,c})$, and the class allocation probabilities $\pi_{nc}$ can be constant across people or depend on characteristics of person $n$, typically through a multinomial logit class allocation model.

\begin{proposition}[Latent class models]\label{prop:lc}
A latent class model with $C$ classes in preference space, with class-specific coefficients $\theta_c$, none of which has $\beta_{\TC,c}=0$, and class allocation probabilities $\pi_{nc}$, and a latent class model with $C$ classes in WTP space, with class-specific coefficients $\phi_c$ and the same class allocation probabilities, are the same model, with $\hat\phi_c = h(\hat\theta_c)$ and identical log-likelihoods. Standard errors of the class-specific WTP measures obtained in WTP space are identical to Delta method standard errors from preference space.
\end{proposition}

The proof is immediate. Each class is a model with non-random coefficients, so the equivalence of the two spaces applies within each class, and hence also to the mixture, in line with Principle~\ref{pr:equiv}. Formally, moving from preference space to WTP space replaces each $\theta_c$ by $h(\theta_c)$ and leaves the class allocation parameters unchanged, which is a one-to-one reparameterisation of a finite-dimensional parameter vector (as no class has $\beta_{\TC,c}=0$), so Proposition~\ref{prop:fixed} applies directly. The proviso matters in practice. A class that ignores cost, as in models of cost non-attendance, can be represented in preference space but not in WTP space, and a class with a cost coefficient close to zero leads to very large valuations that are poorly identified in WTP space, even though the model, and hence the implied WTP, is the same as in preference space. Any difference in fit between ``a latent class model in preference space'' and ``a latent class model in WTP space'' with the same number of classes and the same allocation model indicates convergence to different local optima, which is common with latent class likelihoods \citep{HessBunch2026} and is made more likely by the non-linearity of the WTP-space utility.

Equivalence is lost when restrictions are imposed across classes in the form that is natural in each space, as these restrictions apply to different quantities. Restricting the time coefficient to be the same in all classes in preference space, $\beta_{\TT,c}=\beta_{\TT}$, implies class-specific VTTs that are inversely proportional to the cost coefficients, $\VTT_c = \beta_{\TT}/\beta_{\TC,c}$. Restricting the VTT to be the same in all classes in WTP space, $\beta_{v,\TT,c} = \beta_{v,\TT}$, instead implies that the time coefficients are proportional to the cost coefficients, $\beta_{\TT,c} = \beta_{\TC,c}\beta_{v,\TT}$, so that, if all valuations are restricted in this way, the classes differ only in the scale of monetised utility. These are different models. Either restriction can be imposed in either space, as a simple equality in one and as a non-linear restriction in the other, so a difference in fit between them compares two restrictions, not two spaces. A common cost coefficient, by contrast, takes the same form in both spaces, as the cost coefficient is the same in both. Similarly, with deterministic heterogeneity within classes, restricting interaction parameters to be the same across classes is a restriction on a particular quantity, unless it applies to cost, such as a common income elasticity $\lambda$.

\subsubsection{Discrete mixtures with separate support for each coefficient}\label{sec:dm}

A second specification gives each random coefficient its own discrete distribution, with its own support points and probabilities, and treats the coefficients as independent \citep[e.g.][]{HessBierlairePolak2007}. With $R$ support points for the cost coefficient and $M$ for the time coefficient, or for the VTT in WTP space, this is a latent class model with $R\times M$ classes whose probabilities are products of the probabilities of the support points, but in which the support points are shared across classes. In preference space, the partial derivative of utility with respect to time then takes only $M$ distinct values across the $R\times M$ classes, while the VTT, as the ratio $\beta_{\TT,m}/\beta_{\TC,r}$, in general takes $R\times M$. In WTP space, it is the other way round: the VTT takes only $M$ values, while the partial derivative with respect to time, $\beta_{\TC,r}\beta_{v,\TT,m}$, in general takes $R\times M$. The partial derivatives cannot be matched and, by Principle~\ref{pr:equiv}, the two specifications are different models. They differ both in the assumption of independence (Section~\ref{sec:corr}) and in which values are shared across classes. Free class probabilities $\pi_{rm}$ remove the first difference but not the second, and only when each class is also given its own coefficients do both become the same unrestricted latent class model, which, by Proposition~\ref{prop:lc}, is the same in the two spaces. The same applies to non-parametric approximations that use a large, fixed grid of support points, such as logit mixed logit \citep{Train2016}, as a regular grid in one space does not map onto a regular grid in the other.

\subsection{Continuous mixtures}\label{sec:continuous}

In a continuous mixture, the analyst chooses parametric distributions for the random parameters. In preference space, these are the coefficients, which, with utility linear in the attributes, are the partial derivatives of utility with respect to the attributes, with density $f(\theta\mid\Omega)$ in \eqref{eq:mix_cont}. In WTP space, this is the case only for the cost coefficient. For the other attributes, the distributions are instead chosen for the valuation parameters, with density $f^\phi(\phi\mid\Omega^\phi)$, and the partial derivatives of utility with respect to these attributes are the products $\beta_{\TC}\beta_{v,\TT}$ and $\beta_{\TC}\beta_{v,\CH}$, whose distributions follow from those chosen for their two components. These implied distributions in general do not belong to the families that would be chosen in preference space, and the two models coincide exactly only when they do (Principle~\ref{pr:equiv}). When they do not, the difference in log-likelihood is a comparison of one distributional assumption with another, for the same underlying utility function. An exception is a non-random cost coefficient. The link between the two spaces is then linear, $\beta_{k,n} = \beta_{\TC}\beta_{v,k,n}$, so that a valuation with, for example, a normal or lognormal distribution corresponds to a coefficient with a distribution of the same form, and the two spaces are equivalent. This is, however, likely to be a poor specification in either space, as it rules out heterogeneity in cost sensitivity, which is typically substantial, as well as any scale heterogeneity, and attributes all heterogeneity in WTP to the non-cost attributes.

For example, a preference-space model specifies a distribution for $\beta_{\TT}$, and the VTT then follows as the ratio $\beta_{\TT}/\beta_{\TC}$, whose distribution is generally of a different form\footnote{Its moments may not exist if the density of the cost coefficient does not fall to zero quickly enough near zero, e.g.\ for normal, uniform or triangular distributions whose support reaches zero \citep{DalyHessTrain2012}, and it can then take extreme values \citep{SillanoOrtuzar2005}.}. \citet{ScarpaThieneTrain2008} illustrate the general case, specifying a negative lognormal cost coefficient together with normal distributions for the other coefficients in preference space, or for the valuations in WTP space, both without and with correlation. As normal distributions allow both signs, Proposition~\ref{prop:ln} does not apply, and their better fit in WTP space, in both cases, reflects the different distributional families. In the Bayesian comparison of \citet{SonnierEtAl2007}, the gains from the WTP parameterisation come from being able to place a sensible prior directly on WTP, which is again a distributional choice. Neither space is the ``right'' one, and the implied distributions of both marginal utilities \emph{and} WTP should be reported. WTP space also makes it convenient to test restrictions on the WTP distribution directly \citep{ThieneScarpa2009}, but the same tests are available in preference space using Delta method standard errors for the moments of the implied WTP distribution, where these moments exist \citep[][Appendix 2]{DalyHessDeJong2012,ScacciaEtAl2023,DalyHessOrtuzar2023}.

When all coefficients have a fixed sign, taking logs of the absolute values of $\beta_{\TT,n} = \beta_{\TC,n}\beta_{v,\TT,n}$ and $\beta_{\CH,n} = \beta_{\TC,n}\beta_{v,\CH,n}$ turns the products into sums,
\begin{equation}\label{eq:logid}
\ln|\beta_{\TT,n}| = \ln|\beta_{\TC,n}| + \ln|\beta_{v,\TT,n}|,\qquad \ln|\beta_{\CH,n}| = \ln|\beta_{\TC,n}| + \ln|\beta_{v,\CH,n}|,
\end{equation}
where the left-hand sides are preference-space coefficients and the right-hand sides express them in terms of the WTP-space parameters, with the cost coefficient common to both. In logs, the link between the two spaces is therefore linear, which leads to the following result.

\begin{proposition}[Exact equivalence with unrestricted correlation of log-coefficients]\label{prop:ln}
Suppose that $\beta_{\TC}<0$ and that each non-cost coefficient has a fixed sign, so that each valuation parameter $\beta_{v,k} = \beta_k/\beta_{\TC}$ has the opposite sign to the corresponding preference-space coefficient $\beta_k$. Let $\mathbf{a}_n = (\ln|\beta_{\TC,n}|,\ln|\beta_{\TT,n}|,\ln|\beta_{\CH,n}|)^\top$ and $\mathbf{b}_n = (\ln|\beta_{\TC,n}|,\ln|\beta_{v,\TT,n}|,\ln|\beta_{v,\CH,n}|)^\top$. Stacking \eqref{eq:logid}, together with the fact that the cost coefficient is the same in both spaces, gives $\mathbf{a}_n = A\mathbf{b}_n$ with
\begin{equation}\label{eq:Amat}
A = \begin{pmatrix}1&0&0\\1&1&0\\1&0&1\end{pmatrix}.
\end{equation}
If, in WTP space, $\mathbf{b}_n = \mu_b + L_b\boldsymbol{\eta}_n$, with $L_b$ an unrestricted lower triangular matrix, cost first in the ordering, and $\boldsymbol{\eta}_n$ a vector of independent draws from any given distribution, then $\mathbf{a}_n = A\mu_b + AL_b\boldsymbol{\eta}_n$, where $AL_b$ is again an unrestricted lower triangular matrix. A preference-space model and a WTP-space model of this form, with the same distribution for $\boldsymbol{\eta}_n$, are therefore the same model, and with the same draws, the two simulated log-likelihoods are identical draw by draw, not just asymptotically.
\end{proposition}

In terms of Principle~\ref{pr:equiv}, the joint distribution of the partial derivatives of utility is the same in both spaces. The proof is immediate from \eqref{eq:logid} and the fact that, as $A$ is itself lower triangular and invertible, $AL_b$ ranges over all lower triangular matrices as $L_b$ does\footnote{The ordering condition matters in practice. In Stata, \texttt{mixlogitwtp} \citep{Hole2016} with the \texttt{corr} option and lognormal valuations estimates the same model as a correlated-lognormal \texttt{mixlogit} in preference space \citep{Hole2007}, but places price last, so the two give slightly different simulated log-likelihoods for a finite number of draws, which is simulation noise, not evidence about the two spaces.}. The proposition generalises directly to any number of non-cost attributes. Constants included in monetised utility are covered by treating them as further non-cost attributes.

The most important case is that of normal draws $\boldsymbol{\eta}_n$, which gives jointly lognormal coefficients with an unrestricted covariance matrix in both spaces. A preference-space model with a negative lognormal cost coefficient and, for example, a negative lognormal time coefficient is then the same model as a WTP-space model with a negative lognormal cost coefficient and a positive lognormal VTT, provided the full covariance matrix is estimated in both\footnote{The same holds for other distributions of $\boldsymbol{\eta}_n$, such as uniform or Student $t$, but only the first coefficient in the ordering then follows the corresponding log-distribution, as the logs of the others are weighted sums of independent draws, which, unlike for normals, are generally of a different form. Independent loguniforms in the two spaces therefore give different models, as do independent lognormals (Section~\ref{sec:corr}).}. For an attribute with a positive coefficient, the valuation then needs a negative lognormal.

Proposition~\ref{prop:ln} makes the central point concrete. When a lognormal WTP-space model fits worse (or better) than a lognormal preference-space model, beyond simulation noise and local optima, at least one of them must have been estimated with a restricted covariance matrix, typically both with diagonal ones, and the two restrictions are then \emph{different}. Independence in WTP space assumes that people who are more sensitive to cost are also more sensitive to time, while independence in preference space assumes that more cost-sensitive people have a lower VTT (Section~\ref{sec:corr}), and if the assumption is wrong, the implied VTT distribution is biased. Neither is a neutral assumption, and the difference in fit tells us which restriction the data prefer, not which ``space'' is better, consistent with the finding of \citet{HoleKolstad2012} that allowing for correlation matters more for fit than the choice of space.

\subsection{Independence and correlation}\label{sec:corr}

The reason independence is not neutral is that it is only defined relative to a parameterisation. \citet{TrainWeeks2005} and \citet{ScarpaThieneTrain2008} make this point verbally, noting that uncorrelated utility coefficients imply WTPs that are ``correlated in a particular way'', and vice versa, and \citet{HessTrain2017} note that, when the valuation parameters are independent of each other and of the cost coefficient, the only correlation between the partial derivatives of utility is that induced by the common cost coefficient, which acts like scale heterogeneity. Here we make the implied correlations explicit. We consider the common case in which the coefficients are specified as mutually independent in one of the two spaces, and derive the correlations that this implies in the other. The results in \eqref{eq:cov_tc}--\eqref{eq:corr_tc}, for independence in WTP space, do not depend on the form of the distributions, and so apply equally to discrete mixtures with separate support for each coefficient (Section~\ref{sec:dm}). For independence in preference space, the VTT is a ratio whose moments may not exist in general, and we give the corresponding result for lognormal distributions below. Estimating only some correlations, with the others fixed to zero in one space, still imposes particular correlations in the other.

Assume, as is common in WTP-space applications, that $\beta_{\TC}$, $\beta_{v,\TT}$ and $\beta_{v,\CH}$ are mutually independent, with means $\mu_{\TC},\mu_{v,\TT},\mu_{v,\CH}$ and variances $\sigma_{\TC}^2,\sigma_{v,\TT}^2,\sigma_{v,\CH}^2$. This is, for example, the default in \texttt{mixlogitwtp} unless the \texttt{corr} option is specified \citep{Hole2016}, so it is likely to underlie much applied work. The implied marginal utilities $\beta_{\TT} = \beta_{\TC}\beta_{v,\TT}$ and $\beta_{\CH} = \beta_{\TC}\beta_{v,\CH}$ are then correlated, whatever the distributions:
\begin{align}
\mathrm{Cov}(\beta_{\TT},\beta_{\TC}) &= \mu_{v,\TT}\,\sigma_{\TC}^2, \label{eq:cov_tc}\\
\mathrm{Cov}(\beta_{\TT},\beta_{\CH}) &= \mu_{v,\TT}\,\mu_{v,\CH}\,\sigma_{\TC}^2, \label{eq:cov_tx}\\
\mathrm{Corr}(\beta_{\TT},\beta_{\TC}) &= \frac{\mu_{v,\TT}\,\sigma_{\TC}}{\sqrt{\sigma_{\TC}^2\sigma_{v,\TT}^2 + \mu_{\TC}^2\sigma_{v,\TT}^2 + \mu_{v,\TT}^2\sigma_{\TC}^2}}.\label{eq:corr_tc}
\end{align}
With a positive VTT, people who are more sensitive to cost are therefore \emph{assumed} to be more sensitive to time as well, and to every other monetised attribute. As $\sigma_{v,\TT}\to 0$ the correlation tends to one, and all heterogeneity in $\beta_{\TC}$ becomes heterogeneity in the overall scale of monetised utility. This limiting case arises whenever a valuation is specified as non-random alongside a random cost coefficient, which then implies that the corresponding marginal utility varies in exact proportion to the cost coefficient.

This structure also helps to explain the numerical difficulties reported with WTP-space models with a lognormal cost coefficient \citep{Helveston2023}, at least when the valuations are independent of the cost coefficient or non-random. Draws in the upper tail of $-\beta_{\TC}$ then multiply all utility differences by a large factor, making the simulated log-likelihood steep in those regions and results sensitive to the number of draws. This is a property of the model, not of the space, as the same specification written in preference-space form, unlike a preference-space model with independent coefficients, has exactly the same likelihood.

The mirror image holds in preference space. With independent lognormal $\beta_{\TC}$ and $\beta_{\TT}$, $\ln\VTT = \ln|\beta_{\TT}| - \ln|\beta_{\TC}|$, so that $\mathrm{Corr}(\ln\VTT,\ln|\beta_{\TC}|) = -s_{\TC}/\sqrt{s_{\TC}^2+s_{\TT}^2} < 0$, where $s_{\TC}^2$ and $s_{\TT}^2$ are the variances of the logs, and the logs of all WTP measures are positively correlated with each other through their common denominator.

These opposite assumptions explain the often large differences in fit and in the implied mean and variance of the VTT between such models, with the preference-space version typically having a much longer right tail, driven by small values of $|\beta_{\TC}|$ \citep[e.g.][]{TrainWeeks2005,ScarpaThieneTrain2008}. By Proposition~\ref{prop:ln}, estimating the full covariance matrix removes the difference in the lognormal case.

\subsection{Combining discrete and continuous mixtures}\label{sec:lcmix}

The equivalence of latent class models in Section~\ref{sec:lc} relies on the coefficients being non-random within each class. With continuous mixing within classes \citep[e.g.][]{GreeneHensher2013}, each class is itself a continuous mixture, so the two spaces are equivalent only if the within-class distributions are, as with full-covariance lognormals in every class (Proposition~\ref{prop:ln}).

\section{Conclusions}

This note has sought to address a common confusion in the literature by stressing that WTP space is a reparameterisation of utility, not a different model. By Principle~\ref{pr:equiv}, the two are the same model if and only if, up to the normalisations needed for identification and with the same specification of the error term, they imply the same constants and partial derivatives of utility for every person, or, with random heterogeneity, the same joint distribution of these. With non-random coefficients, this means that the two spaces give identical log-likelihoods, estimates and standard errors, so that any difference in applied work points to non-convergence or to a different specification, and that the cost coefficient in WTP space is estimated, not ``constrained to one'', even though heterogeneity placed on it acts like scale heterogeneity.

With deterministic heterogeneity or non-linearity, the main practical attraction of WTP space, namely that the estimated parameters are the WTP measures, is more limited than often assumed. Covariate effects on the marginal utility of income belong on cost within monetised utility, where they affect all WTP measures, and they, like any non-linearity in cost or in the other attributes, then require WTP to be computed as a ratio of partial derivatives. Centring covariates or normalising non-linear transformations at a reference point retains a single directly estimated WTP, but only at the reference values, and the mean WTP in the sample again requires integrating ratios of partial derivatives over the distribution of the covariates and attribute levels.

With random heterogeneity, differences between the spaces in fit and in the implied WTP distributions reflect differences in the distributions specified, not in the space. Latent class models are the same model in the two spaces, provided no class has a zero cost coefficient, but restrictions across classes on quantities other than cost, discrete mixtures with separate support for each coefficient, and continuous mixtures in general are not. Independence in particular is not a neutral assumption, as it implies opposite correlation structures in the two spaces.

Taken together, WTP can be read directly from the estimates without changing the model only in restrictive cases. Cost and the other attributes must enter monetised utility linearly, with any deterministic heterogeneity entering only through the valuation parameters, and the model must in addition either have no random heterogeneity, be a latent class model with fully class-specific coefficients, or be a continuous mixture with jointly log-distributed coefficients and a full covariance matrix or with a non-random cost coefficient, which is, however, a very restrictive specification. In all other cases, either WTP must still be computed as a ratio of partial derivatives, or the WTP-space model is not equivalent to a preference-space model, and its WTP distribution then reflects a different distributional assumption rather than a more direct estimate. The choice of space is thus a choice of where to place distributional assumptions and restrictions, not a choice between rival models, and analysts should report the implied distributions of both the partial derivatives of utility and the WTP measures.

\section*{Acknowledgements}

The author acknowledges financial support from the European Research Council through the advanced grant 101020940-SYNERGY, and thanks Thijs Dekker for comments on an earlier draft.

\bibliographystyle{apalike}
\bibliography{refs}

@incollection{TrainWeeks2005,
  author    = {Train, Kenneth and Weeks, Melvyn},
  title     = {Discrete choice models in preference space and willingness-to-pay space},
  booktitle = {Applications of Simulation Methods in Environmental and Resource Economics},
  editor    = {Alberini, Anna and Scarpa, Riccardo},
  publisher = {Springer},
  address   = {Dordrecht},
  year      = {2005},
  pages     = {1--16}
}

@article{ScarpaThieneTrain2008,
  author  = {Scarpa, Riccardo and Thiene, Mara and Train, Kenneth},
  title   = {Utility in willingness to pay space: a tool to address confounding random scale effects in destination choice to the {A}lps},
  journal = {American Journal of Agricultural Economics},
  volume  = {90}, number = {4}, pages = {994--1010}, year = {2008}
}

@article{SonnierEtAl2007,
  author  = {Sonnier, Garrett and Ainslie, Andrew and Otter, Thomas},
  title   = {Heterogeneity distributions of willingness-to-pay in choice models},
  journal = {Quantitative Marketing and Economics},
  volume  = {5}, number = {3}, pages = {313--331}, year = {2007}
}

@article{HensherGreene2011,
  author  = {Hensher, David A. and Greene, William H.},
  title   = {Valuation of travel time savings in {WTP} and preference space in the presence of taste and scale heterogeneity},
  journal = {Journal of Transport Economics and Policy},
  volume  = {45}, number = {3}, pages = {505--525}, year = {2011}
}

@article{HoleKolstad2012,
  author  = {Hole, Arne Risa and Kolstad, Julie Riise},
  title   = {Mixed logit estimation of willingness to pay distributions: a comparison of models in preference and {WTP} space using data from a health-related choice experiment},
  journal = {Empirical Economics},
  volume  = {42}, number = {2}, pages = {445--469}, year = {2012}
}

@article{DalyHessTrain2012,
  author  = {Daly, Andrew and Hess, Stephane and Train, Kenneth},
  title   = {Assuring finite moments for willingness to pay in random coefficient models},
  journal = {Transportation},
  volume  = {39}, number = {1}, pages = {19--31}, year = {2012}
}

@article{DalyHessDeJong2012,
  author  = {Daly, Andrew and Hess, Stephane and de Jong, Gerard},
  title   = {Calculating errors for measures derived from choice modelling estimates},
  journal = {Transportation Research Part B},
  volume  = {46}, number = {2}, pages = {333--341}, year = {2012}
}

@article{HessRose2012,
  author  = {Hess, Stephane and Rose, John M.},
  title   = {Can scale and coefficient heterogeneity be separated in random coefficients models?},
  journal = {Transportation},
  volume  = {39}, number = {6}, pages = {1225--1239}, year = {2012}
}

@article{HessTrain2017,
  author  = {Hess, Stephane and Train, Kenneth},
  title   = {Correlation and scale in mixed logit models},
  journal = {Journal of Choice Modelling},
  volume  = {23}, pages = {1--8}, year = {2017}
}

@article{FiebigEtAl2010,
  author  = {Fiebig, Denzil G. and Keane, Michael P. and Louviere, Jordan and Wasi, Nada},
  title   = {The generalized multinomial logit model: accounting for scale and coefficient heterogeneity},
  journal = {Marketing Science},
  volume  = {29}, number = {3}, pages = {393--421}, year = {2010}
}

@article{HessEtAl2017,
  author  = {Hess, Stephane and Daly, Andrew and Dekker, Thijs and Ojeda Cabral, Manuel and Batley, Richard},
  title   = {A framework for capturing heterogeneity, heteroskedasticity, non-linearity, reference dependence and design artefacts in value of time research},
  journal = {Transportation Research Part B},
  volume  = {96}, pages = {126--149}, year = {2017}
}

@article{RoseMasiero2010,
  author  = {Rose, John M. and Masiero, Lorenzo},
  title   = {A comparison of the impacts of aspects of prospect theory on {WTP/WTA} estimated in preference and {WTP/WTA} space},
  journal = {European Journal of Transport and Infrastructure Research},
  volume  = {10}, number = {4}, pages = {330--346}, year = {2010}
}

@article{ScacciaEtAl2023,
  author  = {Scaccia, Luisa and Marcucci, Edoardo and Gatta, Valerio},
  title   = {Prediction and confidence intervals of willingness-to-pay for mixed logit models},
  journal = {Transportation Research Part B},
  volume  = {167}, pages = {54--78}, year = {2023}
}

@article{Helveston2023,
  author  = {Helveston, John Paul},
  title   = {logitr: fast estimation of multinomial and mixed logit models with preference space and willingness-to-pay space utility parameterizations},
  journal = {Journal of Statistical Software},
  volume  = {105}, number = {10}, pages = {1--37}, year = {2023}
}

@article{BalcombeEtAl2009,
  author  = {Balcombe, Kelvin and Chalak, Ali and Fraser, Iain},
  title   = {Model selection for the mixed logit with {B}ayesian estimation},
  journal = {Journal of Environmental Economics and Management},
  volume  = {57}, number = {2}, pages = {226--237}, year = {2009}
}

@article{OwusuCoffieEtAl2016,
  author  = {Owusu Coffie, Rebecca and Burton, Michael P. and Gibson, Fiona L. and Hailu, Atakelty},
  title   = {Choice of rice production practices in {G}hana: a comparison of willingness to pay and preference space estimates},
  journal = {Journal of Agricultural Economics},
  volume  = {67}, number = {3}, pages = {799--819}, year = {2016}
}

@article{ThieneScarpa2009,
  author  = {Thiene, Mara and Scarpa, Riccardo},
  title   = {Deriving and testing efficient estimates of {WTP} distributions in destination choice models},
  journal = {Environmental and Resource Economics},
  volume  = {44}, pages = {379--395}, year = {2009}
}

@article{LadenburgSkotte2022,
  author  = {Ladenburg, Jacob and Skotte, Maria},
  title   = {Heterogeneity in willingness to pay for the location of offshore wind power development: an application of the willingness to pay space model},
  journal = {Energy},
  volume  = {241}, pages = {122900}, year = {2022}
}

@article{Chen2026,
  author  = {Chen, Xianwen},
  title   = {Comparing generalized multinomial logit model and random parameter logit model in preference space and in willingness-to-pay space},
  journal = {Journal of Agriculture and Food Research},
  volume  = {29}, pages = {102998}, year = {2026}
}

@article{SillanoOrtuzar2005,
  author  = {Sillano, Mauricio and Ort{\'u}zar, Juan de Dios},
  title   = {Willingness-to-pay estimation with mixed logit models: some new evidence},
  journal = {Environment and Planning A},
  volume  = {37}, number = {3}, pages = {525--550}, year = {2005}
}

@article{GuHoleKnox2013,
  author  = {Gu, Yuanyuan and Hole, Arne Risa and Knox, Stephanie},
  title   = {Fitting the generalized multinomial logit model in {S}tata},
  journal = {Stata Journal},
  volume  = {13}, number = {2}, pages = {382--397}, year = {2013}
}

@article{Hole2007,
  author  = {Hole, Arne Risa},
  title   = {Fitting mixed logit models by using maximum simulated likelihood},
  journal = {Stata Journal},
  volume  = {7}, number = {3}, pages = {388--401}, year = {2007}
}

@misc{Hole2016,
  author       = {Hole, Arne Risa},
  title        = {{MIXLOGITWTP}: {S}tata module to estimate mixed logit models in {WTP} space},
  howpublished = {Statistical Software Components, Boston College Department of Economics},
  note         = {Version 1.1.0},
  year         = {2016}
}

@article{BHHH1974,
  author  = {Berndt, Ernst K. and Hall, Bronwyn H. and Hall, Robert E. and Hausman, Jerry A.},
  title   = {Estimation and inference in nonlinear structural models},
  journal = {Annals of Economic and Social Measurement},
  volume  = {3}, number = {4}, pages = {653--665}, year = {1974}
}

@incollection{McFadden1974,
  author    = {McFadden, Daniel},
  title     = {Conditional logit analysis of qualitative choice behavior},
  booktitle = {Frontiers in Econometrics},
  editor    = {Zarembka, Paul},
  publisher = {Academic Press},
  address   = {New York},
  year      = {1974},
  pages     = {105--142}
}

@article{HessPalma2019,
  author  = {Hess, Stephane and Palma, David},
  title   = {Apollo: a flexible, powerful and customisable freeware package for choice model estimation and application},
  journal = {Journal of Choice Modelling},
  volume  = {32}, pages = {100170}, year = {2019}
}

@article{GreeneHensher2003,
  author  = {Greene, William H. and Hensher, David A.},
  title   = {A latent class model for discrete choice analysis: contrasts with mixed logit},
  journal = {Transportation Research Part B},
  volume  = {37}, number = {8}, pages = {681--698}, year = {2003}
}

@article{HessBierlairePolak2007,
  author  = {Hess, Stephane and Bierlaire, Michel and Polak, John W.},
  title   = {A systematic comparison of continuous and discrete mixture models},
  journal = {European Transport},
  volume  = {37}, pages = {35--61}, year = {2007}
}

@article{Train2016,
  author  = {Train, Kenneth},
  title   = {Mixed logit with a flexible mixing distribution},
  journal = {Journal of Choice Modelling},
  volume  = {19}, pages = {40--53}, year = {2016}
}

@article{DalyHessOrtuzar2023,
  author  = {Daly, Andrew and Hess, Stephane and Ort{\'u}zar, Juan de Dios},
  title   = {Estimating willingness-to-pay from discrete choice models: setting the record straight},
  journal = {Transportation Research Part A},
  volume  = {176}, pages = {103828}, year = {2023}
}

@article{ReveltTrain1998,
  author  = {Revelt, David and Train, Kenneth},
  title   = {Mixed logit with repeated choices: households' choices of appliance efficiency level},
  journal = {Review of Economics and Statistics},
  volume  = {80},
  number  = {4},
  pages   = {647--657},
  year    = {1998}
}

@misc{HessBunch2026,
  author        = {Hess, Stephane and Bunch, David S.},
  title         = {Get me out of this hole: identifying and avoiding inferior local optima in choice models},
  year          = {2026},
  eprint        = {2506.02722},
  archivePrefix = {arXiv},
  howpublished  = {arXiv preprint arXiv:2506.02722},
  doi           = {10.48550/arXiv.2506.02722}
}

@article{GreeneHensher2013,
  author  = {Greene, William H. and Hensher, David A.},
  title   = {Revealing additional dimensions of preference heterogeneity in a latent class mixed multinomial logit model},
  journal = {Applied Economics},
  volume  = {45},
  number  = {14},
  pages   = {1897--1902},
  year    = {2013},
  doi     = {10.1080/00036846.2011.650325}
}

@article{SmallRosen1981,
  author  = {Small, Kenneth A. and Rosen, Harvey S.},
  title   = {Applied Welfare Economics with Discrete Choice Models},
  journal = {Econometrica},
  year    = {1981},
  volume  = {49},
  number  = {1},
  pages   = {105--130}
}

@article{DagsvikKarlstrom2005,
  author  = {Dagsvik, John K. and Karlstr{\"o}m, Anders},
  title   = {Compensating Variation and {H}icksian Choice Probabilities in Random Utility Models That Are Nonlinear in Income},
  journal = {Review of Economic Studies},
  year    = {2005},
  volume  = {72},
  number  = {1},
  pages   = {57--76}
}

\end{document}